\documentclass[12pt,aps,floats,showpacs,amssymb,tightenlines]{revtex4}

\usepackage{amsmath}
\usepackage{amsfonts}
\usepackage{amscd}
\usepackage{epsfig}
\usepackage{amssymb}
\usepackage{tabularx}
\def\a{\alpha}

\def\G{\Gamma}

\def\m{\mu}

\def\d{\delta}

\def\c{\nabla}

\def\k{\kappa}
\def\r{\rho}

\begin{document}

\title{Obstructions on the horizon geometry from string theory
 corrections to Einstein
gravity}
\author{Gustavo Dotti and Reinaldo J. Gleiser}
\affiliation{Facultad de Matem\'atica, Astronom\'{\i}a y F\'{\i}sica,
Universidad Nacional de C\'ordoba, Ciudad Universitaria,
(5000) C\'ordoba, Argentina}
\email{gdotti@famaf.unc.edu.ar}

\begin{abstract}
Higher dimensional Einstein gravity in vacuum admits static black
hole solutions with an Einstein manifold of {\em non constant
curvature} as a horizon. This gives a much richer family of static
black holes than in four dimensional GR. However, as we show in
this paper, the Gauss-Bonnet string theory correction to Einstein
gravity poses severe limitations on the geometry of a horizon
Einstein manifold. The additional stringy constraints rule out
most of the known examples of exotic black holes with a horizon of
non constant curvature.
\end{abstract}
\pacs{04.50.+h,04.20.-q,04.70.-s}

\maketitle Higher dimensional black holes have come to play an
important role, not only as a theoretical device to gain insight
on problems in 3+1 gravity, but also because of the intriguing
possibility that they could actually be produced in the next
generation of particle accelerators, provided a large extra
dimensions scenario is correct \cite{bhp}. A rich family of
static, vacuum black hole solutions to Einstein equations in $n+2$
dimensions exists, where the horizon manifold $\Sigma_n$ is not
necessarily of constant curvature, as it may belong to the far
less restricted class of Einstein manifolds \cite{gh}. A natural
question to ask is whether or not these black holes could actually
be produced in high energy scattering processes. In \cite{gh2}
this problem is approached by studying the stability of the exotic
black holes in $n+2$ dimensional Einstein gravity, with emphasis
on the case where the horizon Einstein manifolds are spheres or
product of spheres equipped with the inhomogeneous Einstein
metrics discovered by Bohm \cite{bohm}. In this letter we take a
different approach. Since higher dimensional gravity is motivated
by string theory, we consider the effects of the first order
string correction to Einstein gravity, namely, the Gauss-Bonnet
term
\begin{multline} \label{egb2}
{G_{(2)}}_b{}^a = R_{cb}{}^{de}  R_{de}{}^{ca}  -2 R_d{}^c R_{cb}{}^{da}
-2 R_b{}^c R_c{}^a + R  R_b{}^a
 -\frac{1}{4} \d^a_b \left(
R_{cd}{}^{ef}R_{ef}{}^{cd} - 4 R_c{}^d R_d{}^c + R^2 \right)
\end{multline}
String theory predicts that the vacuum equations for the gravitational field are
\cite{str1}
\begin{equation} \label{egb}
0 = {\cal{G}}_b{}^a \equiv \Lambda {G_{(0)}}_b{}^a + {G_{(1)}}_b{}^a
+ \a {G_{(2)}}_b{}^a,
\end{equation}
where $\a$ is related to the string tension, $\Lambda$ the
cosmological constant,
 ${G_{(0)}}_{ab} = g_{ab}$
the spacetime metric and $ {G_{(1)}}_{ab} = R_{ab} -\frac{1}{2} R g_{ab}$
the Einstein tensor. Additional terms of higher order in the curvature
are possible \cite{str2},
 most probably in the form of higher order Lovelock
tensors \cite{Love,str2}. Since Einstein equations  involve only the Ricci
tensor, it is intuitively reasonable that
 replacing a constant curvature horizon
with an Einstein manifold in a black hole solution may  give a new
solution of the field equations. In contrast, the
 Einstein-Gauss-Bonnet (EGB) term from string theory
exposes the full structure of the Riemann tensor, and, as we will
show below, sets non trivial  conditions on the Weyl
tensor of the horizon manifold.\\

We take the horizon $\Sigma_n$ to be a
 {\em Riemannian} manifold  of dimension $n>2$ with
metric $\bar g_{ij}$ (tensors and connection coefficients on
$\Sigma_n$ will be denoted with an overline; coordinate indices
are from the middle of the alphabet). We assume $\Sigma_n$ is an
{\it Einstein} manifold, i.e., one for which
\begin{equation} \label{Ricc-em}
\bar R_{ij} = \k (n-1) \bar g_{ij}.
\end{equation}
Using (\ref{Ricc-em}) in the identity
$\bar \c ^i (R_{ij}-R g_{ij}/2) = 0$ gives $0 = (n-1)(1-n/2)\bar \c _j \k$,
thus
$\k$  in (\ref{Ricc-em}) must be a constant, since we assumed  $n>2$.
Equation (\ref{Ricc-em}) also implies that
\begin{equation} \label{Riem-em}
\bar R_{ij}{}^{kl} = \bar C_{ij}{}^{kl} + \k \left(
\d_i{}^k \;\d_j{}^l - \d_i{}^l \;\d_j{}^k \right),
\end{equation}
where $\bar C_{ij}{}^{kl}$ is the Weyl tensor.
In the particular case where $\bar C_{ij}{}^{kl}=0$, $\Sigma_n$
is a Riemannian manifold of {\em constant curvature} $\k$.
Since the Weyl tensor is identically zero if $n=3$, there
is no distinction between Einstein manifolds and constant curvature
manifolds in three dimensions. However, for $n>3$, constant curvature manifolds
are just special  cases of Einstein manifolds.\\

 Let ${\cal M}$ be
the two dimensional Lorentzian manifold with line element
\begin{equation}
ds^2 = -f(r) dt^2 + g(r) dr^2.
\end{equation}
We will use letters from the beginning of the alphabet for the coordinates
$r,t,$ and underline tensors and connection coefficients for this manifold.
Note that
\begin{equation} \label{chr1}
\underline \G_{tt}{}^r = \frac{f'}{2g} \; , \;\;
\underline \G_{tr}{}^t = \frac{f'}{2f} \; , \;\;
\underline \G_{rr}{}^r = \frac{g'}{2g} \;,
\end{equation}
and that
\begin{equation} \label{R1}
\underline R _{tr}{}^{tr} = \frac{f'g'f+{f'} ^2 g- 2 f" f g}{4 f^2g^2}
\end{equation}

The space-time is taken to be a warped product of $\Sigma_n$ and
${\cal M}$, with metric
\begin{equation} \label{st}
ds^2 = -f(r) dt^2 +  g(r) dr^2 + r^2 \bar g _{ij} dx^i dx^j.
\end{equation}
In the region of interest, $f>0$ and $\partial/\partial t$ is a
time-like Killing vector, orthogonal to the $t=constant$ slices.
If $f=0$ at some $r=r_0$, there is a Killing horizon $\Sigma_n$
in these space-like slices. \\

The non-vanishing Christoffel symbols  of (\ref{st}) are
\begin{equation}
\G^a_{bc} = \underline{\G} ^a_{bc}, \; \; \G^i_{jk} = \bar \G^i_{jk} \; , \;\;
\G^r_{ij} = - \frac{r}{g} \bar g _{ij}, \;\; \G^i_{jr} = \frac{\d^i_j}{r} \; ,
\end{equation}
and the non-trivial components of the Riemann tensor are
\begin{eqnarray} \nonumber
R _{tr}{}^{tr} &=& \frac{f'g'f+{f'} ^2 g- 2 f'' f g}{4 f^2g^2}\\ \nonumber
R _{ri}{}^{rj} &=& \frac{g'}{2 r g^2} \d_i{}^j\\ \label{Rie-st}
R _{ti}{}^{tj} &=& \frac{- f'}{2 r f g} \d_i{}^j\\ \nonumber
R_{ij}{}^{kl} &=& \frac{\bar C_{ij}{}^{kl}}{r^2} +
\left( \frac{\k g -1}{r^2 g} \right)
 \left( \d_i{}^k \;\d_j{}^l - \d_i{}^l \;\d_j{}^k  \right)
\end{eqnarray}
Thus, the    non-zero Ricci tensor components are
\begin{eqnarray} \nonumber
R_t{}^t &=& \frac{-2 f'' f g +f'^2g+f' g' f}{4f^2g^2} -
 \frac{nf'}{2rfg} \\ \nonumber
R_r{}^r &=& \frac{-2 f'' f g +{f'}^2g+f' g' f}{4f^2g^2} + \frac{ng'}{2rg^2}\\
R_i{}^j &=& \frac{r g' f - r f' g + 2 g f (\k g-1)(n-1)}{2 r^2 g^2 f} \d_i^j \; ,
\label{Ric-st}
\end{eqnarray}
and the Ricci scalar is
\begin{equation} \label{R-st}
R = \frac{2r^2f'(f'g+fg')+4nrf(fg'-f'g)-4r^2fgf''+4ngf^2(\k g-1)(n-1)}{(2rfg)^2}.
\end{equation}
The Einstein tensor $G_{(1)}{}_b{}^a$ is diagonal, with components
\begin{eqnarray}\nonumber
G_{(1)}{}_t{}^t &=& \frac{n(n-1) g (1-\k g)-n r g'}{2 r^2g}\\
\label{g1}
G_{(1)}{}_r{}^r &=& \frac{n r f' - n(n-1)f (\k g-1)}{2 r^2 f g}\\
G_{(1)}{}_i{}^i &=& \frac{-2(n-1)f^2 \left[g(\k g-1)(n-2)+g'
\right]+ fg \left[2 r
 (n-1)f'+2 r^2 f''\right] -r^2f f' g' -r^2g f''}{(r f g) ^2}
 \nonumber
\end{eqnarray}
The Gauss-Bonnet tensor $G_{(2)}{}_a{}^b$
may have non-trivial off diagonal elements, these are
\begin{equation} \label{g2inj}
G_{(2)}{}_i{}^j = \frac{\bar C_{ki}{}^{ln} \bar C_{ln
}{}^{kj}}{r^4} \hspace{1cm} j \neq i
\end{equation}
the diagonal elements of $G_{(2)}{}_a{}^b$ are:
\begin{eqnarray} \nonumber
G_{(2)}{}_t{}^t &=& -\left(\frac{\sum_{kjln}
\bar C_{kj}{}^{ln} \bar C_{ln}{}^{kj}}{4 r^4}\right) -\frac{n(n-1)(n-2)(\k g-1)\left[ g
 (n-3)(\k g-1)+ 2 r g'\right]}{4 r^4 g^3} \\ \nonumber
G_{(2)}{}_r{}^r &=& -\left(\frac{\sum_{kjln}
 \bar C_{kj}{}^{ln} \bar C_{ln}{}^{kj}}{4 r^4}\right)
 -\frac{n(n-1)(n-2)(\k g-1)\left[ f
 (n-3)(\k g-1)- 2 r f'\right]}{4 r^4 g^2 f(r)}
\end{eqnarray}
and
\begin{multline} \label{g2}
G_{(2)}{}_i{}^i = \frac{(n-1)(n-2)}{4 r^4 g^3 f^2} \left\{ -(n-3)(\k g-1)
\left[ g(\k g-1)
(n-4)+ 2 r g' \right] f^2 \right. \\
\left. + f \left[ (2 \k r (n-3) f'+ 2 \k r^2 f'') g^2+ (  f'(-g' \k r^2-2r(n-3))
 - 2 r^2f'' ) g + 3 r^2 g' f' \right]  \right. \\
\left. + r^2 (f')^2 g (1-\k g) \right\}+ \left( \frac{4 \sum_{kln}
\bar C_{ki}{}^{ln}
 \bar C_{ln}{}^{ki}
- \sum_{kjln} \bar C_{kj}{}^{ln} \bar C_{ln}{}^{kj}}{4 r^4} \right)
\end{multline}
From the vacuum EGB equations (\ref{egb}),  $0 =
{\cal{G}}_i{}^i-{\cal{G}}_j{}^j$ for all $i$ and $j$, and $0 =
{\cal{G}}_i{}^j, j \neq i$. Using (\ref{st}) (\ref{g1})
(\ref{g2inj}) and (\ref{g2}) these conditions read
\begin{equation} \label{eq1}
\a \sum_{kln} \bar C_{ki}{}^{ln} \bar C_{ln}{}^{kj} =
\frac{\a}{n}
\left( \sum_{kmln} \bar C_{km}{}^{ln} \bar C_{ln}{}^{km}\right) \d_i{}^j
\equiv \a \theta  \d_i{}^j
\end{equation}
From $0 = {\cal{G}}_t{}^t-{\cal{G}}_r{}^r \propto f'g+fg'$ and
(\ref{st}) (\ref{g1}) and (\ref{g2}),  we get  $f = c/g$. We may
then set the constant $c=1$ by rescaling $t$. Introducing
\begin{equation}
f(r) = \k - r^2 \psi(r),
\end{equation}
we find that the remaining equations admit a solution
if  $\theta$ in (\ref{eq1}) is a constant and $\psi(r)$ satisfies
\begin{equation}
\frac{1}{r^n} \left[ r^{n+1} P(\psi(r)) \right]' + \frac{\alpha \theta}{4 r^4} = 0,
\end{equation}
where
\begin{equation}\label{p}
 P(\psi(r)) \equiv \frac{\alpha n(n-1)(n-2)}{4} \psi(r)^2 + \frac{n}{2} \psi(r) - \frac{\Lambda}{n+1}
\end{equation}
In conclusion, the EGB vacuum equations  are:
\begin{equation} \label{c2}
\a \sum_{klm} \bar C_{ki}{}^{lm} \bar C_{lm}{}^{kj} = \a \theta  \d_i{}^j ,
\hspace{.5cm} \theta \text{ constant,}
\end{equation}
\begin{equation} \label{f1}
g(r)^{-1} = f(r) = \k -r^2 \psi(r)
\end{equation}
\begin{equation} \label{f2}
P(\psi(r)) \equiv
\frac{\alpha n(n-1)(n-2)}{4} \psi(r)^2 + \frac{n}{2} \psi(r) - \frac{\Lambda}{n+1} = \frac{\m}{r^{n+1}} - \frac{\a \theta}{4 (n-3) r^4}
\end{equation}
where $\mu$ is an integration constant. If the horizon manifold
has constant curvature (\ref{c2}) is trivial, $\theta=0$ and
(\ref{f1})-(\ref{f2}) reduce to the equations leading to well
known black holes  \cite{w1,w2,bd,w3,atz,chm}. If we drop the
string correction by setting $\a=0$, (\ref{c2}) is trivially
satisfied and we recover the family of solutions whose stability is studied in \cite{gh,ki}.\\

The main result of this letter is equation (\ref{c2}), which sets
the condition imposed by string theory on a candidate Einstein
horizon manifold. Equation (\ref{c2}) poses a severe constraint on
the geometry of the Einstein manifold that rules out most  non
trivial (i.e. non constant curvature) Einstein manifolds. Note
that (\ref{c2}) is both an algebraic and a differential
constraint, since $\c_j \theta=0$. The algebraic constraint is
always satisfied if $n=4$, namely, all four dimensional Einstein
manifolds satisfy an equation like (\ref{c2}) with a non-constant
$\theta$ \cite{knt}. In higher dimensions, however, $\sum_{klm}
\bar C_{ki}{}^{lm} \bar C_{lm}{}^{kj}$ need not be
proportional to $\d_i{}^j$.\\

As an example, we will apply equation (\ref{c2}) to the Bohm
metrics in \cite{bohm,gh2}. We should mention here that
black-holes with Bohm horizons were found to be unstable under
tensor mode perturbations in Einstein gravity \cite{gh2}.\\

The Bohm metrics have positive curvature and are locally given by
\cite{gh2}
\begin{equation} \label{b1}
ds^2 = d \rho^2 + a(\rho)^2 d \Omega_p^2 + b(\r)^2 d \Omega_q^2,
\end{equation}
where $d \Omega_m^2$ is the line element of a unit $m-$sphere. These
can be extended onto manifolds of topology $S^{p+q+1}$ or $S^{p+1}
\times S^q$, as long as
\begin{equation} \label{ic}
a(0)=0, \;\;\; \dot a(0)=1, \;\;\; b(0)= b_o, \;\;\; \dot b(0)=1.
\end{equation}
There are  infinitely many Bohm metrics on $S^{p+q+1}$
corresponding to different choices of $a(\r)$ and $b(\r)$. These
are labeled $Bohm(p,q)_{2m}, m=0,1,2,...$ in \cite{gh2}. There is
also an infinite family on $S^{p+1} \times S^q$, labeled
$Bohm(p,q)_{2m+1}, m=0,1,2,...$ The variable $\rho$ runs from zero
to a value $\rho_f$, and $0 < a(\rho), b(\rho)$ if $0 < \rho_f$
\cite{gh2}.
 Using the results in \cite{gh2}, and introducing
\begin{equation} \label{xyz}
X_a := \frac{\ddot a}{a}+\k, \;\;\;Y_a := \frac{\dot{a}^2-1}{a^2}+\k,
\;\;\; Z_{ab} := \frac{\dot a \dot b}{a b}+\k
\end{equation}
(and analogous definitions for  $X_b$ and $Y_b$), we can write  the
conditions for (\ref{b1}) to  satisfy (\ref{Ricc-em}) as \cite{gh2}
\begin{eqnarray} \nonumber
X_a + q Z_{ab} + (p-1) Y_a  = 0 \\ \label{be}
X_b + p Z_{ab} + (q-1) Y_b  = 0\\ \nonumber
pX_a+qX_b = 0.
\end{eqnarray}
If we further impose (\ref{c2}) we get three more equations
\begin{eqnarray} \nonumber
 p {X_a}^2 +  q {X_b}^2 &=& \theta/2 \\ \label{bc2}
{X_a}^2 + q {Z_{ab}}^2+ (p-1) {Y_a}^2 &=& \theta/2 \\ \nonumber
{X_b}^2 + p {Z_{ab}}^2+ (q-1) {Y_b}^2 &=& \theta/2
\end{eqnarray}
Fixing the conformal factor such that $\k=1$, and  regarding
(\ref{be})-(\ref{bc2}) as algebraic equations on
$X_a,X_b,Y_a,Y_b,Z_{ab}, p, q$ and $\theta$, we find a number of
solutions, many of which are trivial because they have $\theta=0$
and thus correspond to a null Weyl tensor. Inserting the remaining
(algebraic) solutions  in (\ref{xyz}) leaves a unique possibility:
\begin{equation}
p=q-1,\;\; \theta= \frac{2q(2q-1)}{q-1},\;\;
a(\r)=\sqrt{\frac{q-1}{2q-1}}\; \sin \left(
\sqrt{\frac{2q-1}{q-1}}r \right),\;\;
b(\r)=\sqrt{\frac{q-1}{2q-1}}.
\end{equation}
This can easily be recognized as the standard metric on $S^q
\times S^q$, a well known homogeneous Einstein metric which
corresponds to the particular case $Bohm(q-1,q)_1$ in the notation
of \cite{gh2}. Of the countably infinite set of Bohm metrics, only
this one is admissible as a horizon. In particular, no static
black hole in odd spacetime dimensions admits a Bohm horizon.

\section*{Acknowledgments}

This work was supported in part by grants of the Universidad
Nacional de C\'ordoba, Agencia C\'ordoba Ciencia (Argentina),
and by grant NSF-INT-0204937 of the National Science Foundation of the US. The
authors are supported by CONICET (Argentina).

\end{document}